\documentstyle[12pt,fleqn,epsfig]{article}
\def\beq {\begin{eqnarray}}
\def\eeq {\end{eqnarray}}
\def\beqn {\begin{eqnarray*}}
\def\eeqn {\end{eqnarray*}}

\def\ra {\rightarrow}
\def\etal {{\it et al}.}

\def\bra {\langle}
\def\ket {\rangle}
\def\GeV{\mbox{GeV}}
\def\MeV{\mbox{MeV}}

\def\ni {\noindent}

\newcommand{\AmS}{{\protect\the\textfont2
  A\kern-.1667em\lower.5ex\hbox{M}\kern-.125emS}}

\begin{document}

\baselineskip 0.9cm


\begin{center}
{\Large Structure of the constituent quark
and \\quark distributions of
hadrons\footnote{Talk presented at International Conference 'Quark Lepton
Nuclear Physics (QULEN97)', May 20-23, 1997, Osaka, Japan}}

\vspace{1.5cm}

{\large Katsuhiko Suzuki\footnote{Alexander von Humboldt fellow, 
e-mail: ksuzuki@physik.tu-muenchen.de}}\\
{\em Physik-Department, 
Technische Universit\"{a}t M\"{u}nchen, \\
D-85747 Garching, Germany}

\vspace{2cm}

Abstract
\end{center}

\baselineskip 0.75cm

\ni
We study the structure of constituent quarks by dressing bare quarks with
the Goldstone bosons and its implications for quark distribution functions
of hadrons $f_1(x)$, $g_1(x)$ and $h_1(x)$.  
In particular we discuss effects of the dressing on the nucleon spin 
structure, and find that contributions to chiral-odd 
$h_1(x)$ is quite different from those to $g_1 (x)$, which can be measured in
the semi-inclusive polarized deep inelastic scattering.

\vspace{2cm}


\baselineskip 0.75cm

\ni
{\bf 1 Introduction}

Much of the success of the low energy hadron phenomenology is built on the 
constituent
quark picture.  Constituent quarks, which may be understood as quasi-particles
of the non-perturbative QCD vacuum, play an important role to describe hadron
properties.  This success naturally leads us to study the quark distribution
function observed in deep inelastic scattering within the
constituent quark picture.  
In fact, the constituent quark models have been applied to calculate 
the structure function by several groups\cite{Kulagin}.  
For the large Bjorken-$x$ region valence constituent quark picture should 
work well as suggested by the quark counting rule.   
However, the valence quark
description is not enough to explain the intermediate and small-$x$ 
behavior of the structure function.  
In ref.~\cite{Kulagin}, inclusion of the pion dressing and higher mass 
spectator process produces a substantial renormalization and provides a correct
small-$x$ behavior for $f_1(x)$ structure function.  
In this talk, we extend it to the spin dependent twist-2 
structure functions $g_1 (x)$ and $h_1 (x)$\cite{Suzuki1}.

At the scale below 1GeV, the relevant degrees of 
freedom are assumed to be constituent quarks (CQ) and Goldstone (GS) bosons.  
Once the bare quarks are dressed by GS bosons ($\pi$, $K$ and $\eta$), 
we can write the constituent $u$-quark Fock state as
\beq
\left| U \right\rangle  = \sqrt{Z}\left| u_0 \right\rangle + 
a_\pi \left| {d \pi ^+} 
\right\rangle +{{a_\pi } \over 2}\left| {u \pi ^0} \right\rangle +a_K\left| 
{s K^+} \right\rangle +{{a_\eta } \over 6}\left| {u \eta }
 \right\rangle   \,\, ,
\label{u-fock}
\eeq
where $\left| u_0 \right\rangle$ is the bare $u$-quark state and $Z$ the
renormalization constant for the bare state.  Similar expressions can be
written for other quarks.  
Using the SU(3) chiral quark model \cite{CQM}, 
we evaluate corrections to quark distributions illustrated in Fig.1.  
The GS boson fluctuation generates the renormalization of the bare state as 
well as depolarization effects on the CQ spin structure by emitting 
the GS boson in a relative P-wave state.  
Since we use the Infinite Momentum Frame, 
the factorization of the subprocess is automatic.  
The unpolarized $u$-quark distribution in the nucleon 
is written as 
\beq
u(x) = Z u_0(x)+P_{u \pi / d} \otimes d_0 + 
V_{u / \pi }\otimes P_{\pi \, d / u} \otimes u_0+
{1 \over 2}P_{u \, \pi/ u} \otimes u_0  +  \cdots \,\, ,
\label{u_N}
\eeq
where $u_0(x)$ and $d_0(x)$ are bare quark distributions in the
nucleon, and $\otimes$ denotes the convolution integral.  
$P(y)_{j \, \alpha / i}$ is the 
splitting function which gives the probability to find a CQ 
$j$ carrying the light-cone momentum fraction $y$ together with a 
spectator GS boson $(\alpha = \pi,K,\eta)$, both of which coming from 
a parent CQ $i$.  $V_{i / \alpha }$ is the $i$-quark
distribution in the GS boson $\alpha$.  
$P(y)_{j \, \alpha / i}$ is given by
\beq
P(y)_{j \, \alpha / i}={1 \over {8\pi ^2}}\left( {{{g_A \, \bar m} 
\over {f }}} \right)^2\int_{}^{} {dk_T^2}
{{(m_j - m_i y)^2 + k_T^2} \over {y^2 (1-y) 
\left[ {m_i^2-M^2_{j \alpha}} \right]^2}}, 
\label{spil-f}
\eeq
where $m_i, m_j, m_{\alpha}$ are 
the particle masses, 
$M^2_{j \, \alpha } = ({m_j^2+k_T^2})/y 
+ ({m_{\alpha}^2 + k^2_T})/({1-y})$ and $\bar m = (m_i + m_j) / 2$.   
The integral (\ref{spil-f}) requires a 
momentum cutoff to set the scale of the chiral effective theory, and 
we adopt the exponential cutoff, exp[($m_i^2 - M^2_{j \alpha})/ 4 \Lambda
^2]$.    
We use $g_A=1$, $m_u = 360 \MeV$ and $m_s = 570 \MeV$ in the following
calculations.

We also calculate contributions to the spin structure function in terms of
the spin-dependent splitting functions $\Delta P (y)_{j \, \alpha / i}$.  The 
process (b) never contributes to $g_1(x)$, because 
the GS bosons carry no spin.

\begin{figure}[htb]
\begin{minipage}[t]{55mm}
\psfig{file=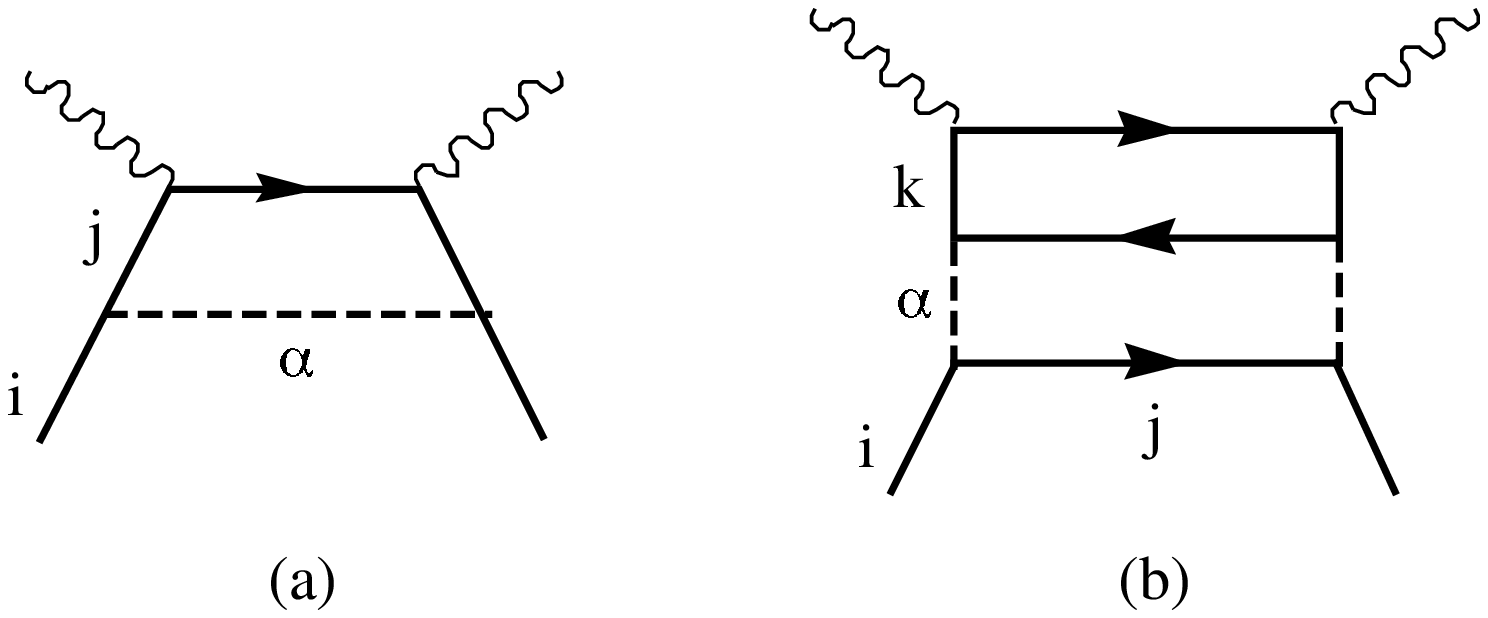,height=0.98in}
\vspace{-0.2cm}
\caption{GS boson dressing}
\label{fig:largenenough}
\end{minipage}
\hspace{\fill}
\begin{minipage}[t]{70mm}
\begin{center}
\psfig{file=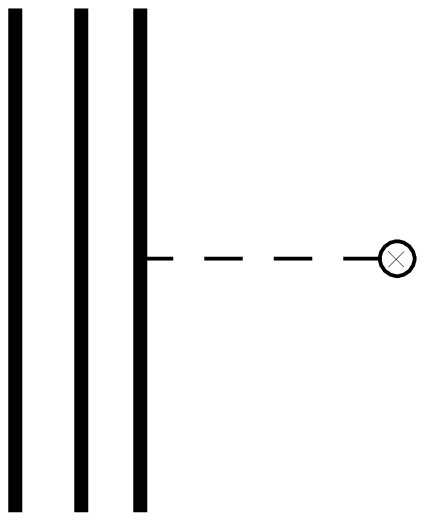,height=0.8in}
\end{center}
\caption{Pion cloud contribution}
\label{fig:toosmall}
\end{minipage}
\end{figure}

\vspace{0.7cm}

\ni
{\bf 2 Gottfried sum and nucleon spin structure in CQM}

Let us discuss the Gottfried sum and the nucleon spin structure in the chiral
constituent quark model, as first done by Eichten {\etal}\cite{Eichten}.  
By using the dressed quark distributions, the Gottfried sum is given by
\beq
GSR = \int_0^1 {\frac{dx}{x}\;}\left[ {F_2^p(x)-F_2^n(x)} \right] 
={1 \over 3} \left( 1 - 2 \left\langle {P_\pi } 
\right\rangle  \right) \; .  
\label{gsr}
\eeq
where $\bra {P_\pi } \ket = \int_0^1 dy  {P_\pi } $, the first
moment of the splitting function.  
We fix the cutoff to reproduce the empirical value of the Gottfried
sum, and find 
GSR $\simeq 0.23$ ($Z=0.67$) with $\Lambda = 1.4 \GeV$.  
This value should be understood as an upper 
bound of the cutoff in this model, based on the assumption that 
the violation of the Gottfried sum is entirely given 
by the GS boson dressing alone.   

We next study the nucleon spin structure.  
The spin fraction of the constituent $u$-quark 
is modified from their bare quark values as follows: 
\beq
\Delta u = Z \Delta u_0 + \frac{1}{2} \left\langle 
 {\Delta P_\pi } \right\rangle \Delta u_0 + 
 \left\langle {\Delta P_\pi } \right\rangle  \Delta d_0 + 
\frac {1}{6} \left\langle {\Delta P_\eta }
\right\rangle  \Delta u_0 \; ,
\label{su}
\eeq
where $\Delta q_0$ are the spin fractions of the bare quarks.  
Similar expressions 
can be written for $d$-quark by replacing $(u,d)$ with $(d,u)$.  
We find that the first moment of the spin-dependent splitting function 
is small and negative, $\bra \Delta P_{\pi} \ket  = -0.06$, which means
depolarization caused by the P-wave coupling to 
the GS bosons is a significant effect.

As a first rough estimate, we start from the naive $SU(6)$ quark model values, 
$\Delta u_0 = {\frac{4}{3}}$, $\Delta d_0 = - \frac{1}{3}$.   
Inserting these values into eq.~(\ref{su}), 
we obtain $\Delta u = 0.86$, $\Delta d = -0.29$, 
$\Delta s = -0.006$, and $\Sigma = 0.56 $.   
%
%
%
%
Considerable part of the CQ spin is transfered to the orbital
angular momentum of the GS bosons, but these values are still far from the 
empirical data.  
If we allow ourselves to vary the momentum cutoff, it would be 
possible to obtain a value around  $\Sigma \sim 0.3$.  
However, the agreement with 
the nucleon axial-vector coupling constant $G_A = \Delta u - \Delta d $ is 
then lost.  
We therefore conclude that the nucleon spin 
problem cannot be solved by the GS boson dressing alone, but the 
depolarization effects are nevertheless significant.

Let us investigate some necessary further steps. 
Up to now we deal only with the $SU(3)$ octet of GS  
bosons to build up the CQ structure, but the constituent quark
already has non-trivial spin structure due to the $U(1)_A$ axial anomaly of
QCD.  Also, the pion cloud directly contributes to the axial vector current  
(Fig.2).  
The relativistic effects of the quark
motion in the nucleon sould be also taken into account.  
Including all these contributions we find in the chiral limit 
($m_\pi = 0$) 
$\Delta u = 0.93$,  $ \Delta d = -0.43$, 
$\Delta s = - 0.12$, and $ \Sigma = 0.21$\cite{Suzuki1}, 
%
%
%
%
which are now reasonably consistent with the empirical values.  
Although we cannot draw strong conclusion from the model dependent analysis, 
the present results combined with other contributions considered above  
go altogether into the right direction of providing a reasonable
description of the nucleon spin structure.

\vspace{1cm}

\ni
{\bf 3 Chiral-odd structure function $h_1(x)$}

We now focus on the chiral-odd transversity  spin structure
function $h_1(x)$.  
The chiral-odd $h_1(x)$ distribution corresponds to a target 
helicity-flip amplitude in the helicity basis, 
and provides a correlation
between the left- and right-handed quarks, which may tell us more about the 
chiral dynamics of QCD.  In the non-relativistic limit $h_1(x) = g_1(x)$, and 
$h_1(x)$ is slightly larger than that
of $g_1(x)$ in the simple relativistic quark model.

We use the same technique to calculate the splitting function 
$\delta P (x) $ for $h_1(x)$ structure function, and find that a relation 
\beq
P(x) + \Delta P(x) = 2 \delta P(x)
\eeq
holds among the GS boson corrections, which implies a saturation of  
Soffer's inequality \cite{Soffer} in the chiral quark model.  
Note that $\delta P  (x)$ is positive in whole $x$ region.  
The first moment of the transversity splitting function is 
$\left\langle {\delta P_\pi } \right\rangle = 0.05$, in contrast to the
negative value of the logitudinal case, $\bra \Delta P_\pi \ket  = -0.06$.

We shall show how the $g_1(x)$ and $h_1(x)$ distributions are modified by the 
GS boson fluctuations.  To get input distributions we use a quark-diquark
model developed previously.   
Regarding the $u$-quark case, 
relative magnitudes of $f_1^u(x)$, $g_1^u (x)$ and $h_1^u (x)$ 
are not modified very much,  though   
the renormalization reduces 
these distribution functions from their original ones.  
The axial charge and tensor charge, first moments of $g_1(x)$ and $h_1(x)$, 
are modified as, ($\Delta u $, $\delta u $) 
$= (1.01, 1.17) \ra (0.65, 0.80)$.  

However, effects of the GS boson dressing are apparent in the 
$d$-quark case.  The $d$-quark tensor charge is much reduced ($\sim 50 \% $),  
whereas corrections to the axial charge are 
small, ($\Delta d $, $\delta d$) = $(-0.25, -0.29) \ra (-0.22, -0.15)$.  
Recent lattice QCD simulation indicates $\Delta u < \delta u $
and $| \Delta d | > | \delta u | $, which is consistent with our results.
We show in Fig.3 $g_1^d(x)$ [a] and $h_1^d(x)$ [b]  with bare results by dashed
curves and dressed cases by the solid ones.  It is clearly seen that
$h_1^d(x)$ is reduced very much.

\vspace{0.3cm}
\begin{figure}[htb]
\begin{center}
\psfig{file=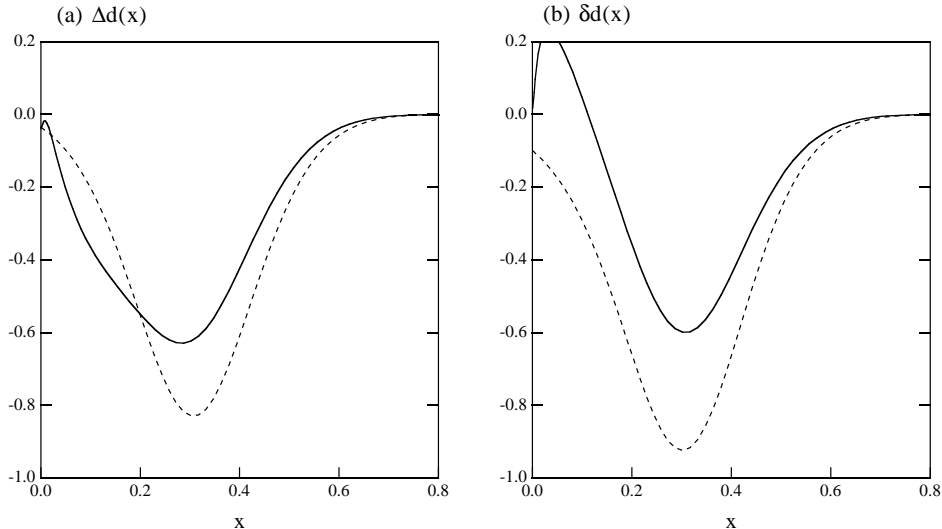,height=2.8in}
\vspace{-0.3cm}
\caption{$g_1(x)$ and $h_1(x)$ with the GS boson dressing[2]}
\label{fig:h1}
\end{center}
\end{figure}

\vspace{0.5cm}

We want to point out that such a difference between $g_1(x)$ and $h_1(x)$ can 
be checked by semi-inclusive polarized deep inelastic scattering.  
We emphasize here the 
semi-inclusive DIS on the transversely polarized nucleon with pion
production.  We can extract 
${h_1^d(x)} / {h_1^u(x)}$ from the angle weighted spin asymmetry of this
process by obserbing $\pi^+$ and $\pi^-$ in the final  
state\cite{Kotz,Suzuki2}.  
We can also obtain ${g_1^d(x)} / {g_1^u(x)}$ using the longitudinally
polarized nucleon.  
We show ${h_1^d(x)} / {h_1^u(x)}$ (solid) and ${g_1^d(x)} / {g_1^u(x)}$
(dashed) in Fig.4\cite{Suzuki2}.   
Due to the existence of the GS boson dressing their shapes are quite 
different.  Simple valence quark model gives almost the same curve for 
both cases.

%
\vspace{0.3cm}

\begin{figure}[htb]
\begin{center}
\psfig{file=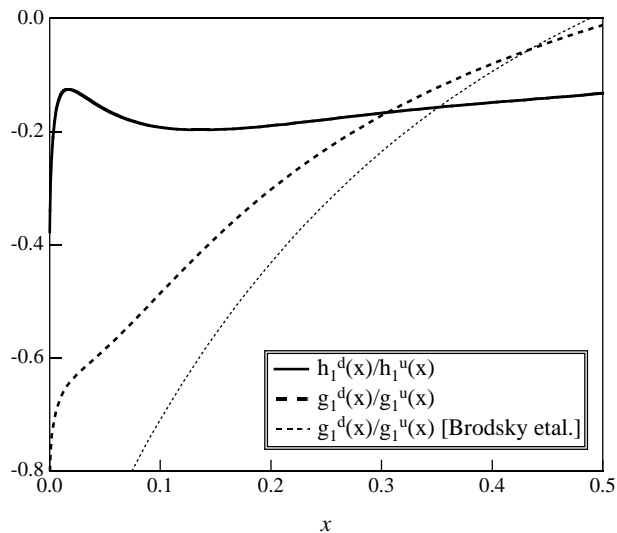,height=2.8in}
\vspace{-0.3cm}
\caption{ ${ h_1^d(x)}/ {h_1^u(x)}$ and ${g_1^d(x)}/ {g_1^u(x)}$ with the 
GS boson dressing[3]}
\label{fig:semi}
\end{center}
\end{figure}

\vspace{1cm}

\ni
{\bf 4 Conclusions}

We have investigated aspects of constituent quark structure in the deep 
inelastic processes by introducing the Goldstone boson fluctuations.   This 
dressing generates the renormalization of the bare CQ state and 
depolarization on the quark spin structure.   
It is also found that this approach gives a large enhancement of the 
sea quark distribution in the pion, whose momentum fraction is twice as large 
as that of the nucleon sea\cite{Suzuki1}.  
Experimental tests of $h_1(x)$ 
distribution function will 
provide new insights into the role of the constituent quarks  
as quasi-particles.  

We would like to thank W. Weise for valuable discussions, careful reading of 
the manuscript and collaboration of ref.~[2].  
This work is supported in part by the Alexander von Humboldt foundation and by
BMBF.

\end{document}